\documentclass{article}

\usepackage{arxiv_custom}

\usepackage[utf8]{inputenc}
\usepackage[T1]{fontenc}
\usepackage{hyperref}
\usepackage{url}
\usepackage{graphicx}
\usepackage{natbib}
\usepackage{float}

\title{NAR db status Version 2 and miRNAverse: Over Two Years of Manual Meta-Registry Curation and Updates}

\author{Marcel Friedrichs\\
	Bioinformatics / Medical Informatics Department\\
        Bielefeld University\\
        Bielefeld, NRW, Germany\\
        \texttt{research@mfriedrichs.me}\\
        \texttt{ORCID: \href{https://orcid.org/0000-0001-9846-7212}{0000-0001-9846-7212}}\\
	\And
	Cassandra Königs\\
	Bioinformatics / Medical Informatics Department\\
        Bielefeld University\\
        Bielefeld, NRW, Germany\\
        \texttt{ORCID: \href{https://orcid.org/0000-0001-8991-9272}{0000-0001-8991-9272}}\\
}

\renewcommand{\shorttitle}{NAR db status Version 2 and miRNAverse}

\hypersetup{
pdftitle={NAR db status Version 2 and miRNAverse},
pdfsubject={q-bio.NC, q-bio.QM},
pdfauthor={Marcel Friedrichs, Cassandra Königs},
}

\begin{document}
\maketitle

\begin{abstract}
Previously, we reported on a new meta-registry for NAR published databases focusing on high-quality annotations regarding database availability and longevity. With over two years of continued manual curation, here, we report on recent updates and additions. Furthermore, the available annotations as well as the underlying database structure have been unified with the miRNAverse meta-registry. This allows for more in-depth insights as well as easier curation and future developments shared across both meta-registries. NAR db status currently provides annotations for 2,082 databases and miRNAverse for 194 databases. With the oldest annotation revision from June 2022 and the newest from January 2025, NAR db status spans two and a half years of continued manual curation. NAR db status is available at \url{https://nardbstatus.de} and miRNAverse at \url{https://mirnaverse.de}.
\end{abstract}



\section*{Introduction}
Databases, datasets, and other sources of information are abundantly available for almost every area of scientific inquiry. New ones are constantly created, existing ones maintained, but, unfortunately, also discontinued or abandoned.

Meta-registries are maintained to make these data sources findable using different criteria and to provide users with insights via annotations \citep{nar2024, nar2025, Ma2022, Ison2019, Pampel2023, FAIRsharing2019} (\url{https://scicrunch.org}, \url{https://integbio.jp/dbcatalog/?lang=en}, \url{https://hsls.pitt.edu/obrc/}),
or to aggregate data from multiple meta-registries such as BiŌkeanós (\url{https://biokeanos.com}).
Others focus on the consistent resolution of data source's identification systems \citep{Hoyt2022Bioregistry, BernalLlinares2020}. 

Previously, we reported on a new meta-registry for NAR-published databases, NAR db status, with a strong focus on longevity information and annotation transparency \citep{Friedrichs2023}. More recently, a separate meta-registry miRNAverse was established, focusing on miRNA databases, species annotations, and database availability \citep{HanselFrse2024}. In the following, we describe recent developments to unify both systems and annotations while maintaining their distinct database subsets. Furthermore, after two years of continuous manual curation, the current state of database longevity is described and compared to previous statistics.



\section*{Materials and Methods}

Following, we describe the process of unifying both meta-registries NAR db status and miRNAverse. This process is split into multiple steps. Afterwards, the implementation of new statistics is described.

\subsection*{Identifiers}
The NAR db status meta-registry used the NAR catalog IDs to reference database entries. As they are defined by the NAR catalog maintainers, it is impossible to reuse and expand them when combining both meta-registries. Furthermore, the NAR catalog IDs are not unique, meaning that multiple IDs represent the same database. These equivalences were previously annotated in NAR db status and used for de-duplication in statistics. As miRNAverse is a registry created from scratch, database entries were assigned unique number IDs.

For unifying both meta-registries, a new ID system is introduced following the schema of a "DB" prefix, followed by an incrementing number. For example, the DrugBank database was assigned the ID "DB94". The NAR catalog and miRNAverse entries IDs are manually linked to these newly created database IDs. This transition further allows for a more precise annotation of databases. NAR catalog entries sometimes refer to multiple databases, for example, entry 163 "BioMuta and BioExpress". Previously, this entry was annotated as a single database. The new ID system allows multiple database IDs to link to the same NAR catalog ID and subsequently the databases to be annotated independently. All NAR catalog and original miRNAverse IDs are automatically resolved to corresponding DB IDs on the website.

\subsection*{Cross-references}
The first versions of the NAR db status and miRNAverse provided cross references to multiple meta-registries. Existing cross-references from both registries have been merged into the newly created database entries. BiŌkeanós has been added as a new meta-registry for cross-referencing, as well as a convenient source of further references. A PHP admin script is able to parse the BiŌkeanós index and extract potential new cross-references from entries with NAR catalog IDs. These potential cross-references are provided to annotators in the respective database entries for quick validation and addition. Overall, cross-references to the following meta-registries are available: \textit{Database Commons}, \textit{FAIRsharing}, \textit{bio.tools}, \textit{re3data}, \textit{SciCrunch}, \textit{Integbio}, \textit{BiŌkeanós}, \textit{OBRC}, \textit{Bioregistry}, and \textit{Identifiers.org}.

\subsection*{Publications}
NAR catalog entries are associated with a specific publication, which was reproduced in the NAR db status entries. In contrast, miRNAverse entries are annotated with all available publications for the specific database. To unify both meta-registries, all combined database entries have been manually annotated with all available publications. The publications' citation counts are regularly and automatically updated using the OpenCitations API \citep{Peroni2020}.

\subsection*{Annotation Revisions}
Previously, miRNAverse entries were annotated without any revision mechanism. All entry's annotations without corresponding NAR db status entry have been manually updated as their initial revision. Subsequent annotation updates are performed in the same revision system as NAR db status. Previously, annotation revisions included database URLs and the NAR publication DOI. With the addition of all available publications and general database entries, URLs and publications have been transferred from the revisions to general entry properties. Where multiple NAR entries have been combined into a single database entry, their revisions have been merged. In case a NAR entry refers to multiple databases, the revisions have been transferred to each database entry and validated with the respective databases. Revisions now represent the availability, bulk download availability, and the last update year of a database at the specific time.

\subsection*{Further Annotation Changes}
Additional annotations from both registries are moved to the new database entries, namely the NAR categories, species, and descriptions. In addition to the database name, also an abbreviation may now be annotated. Finally, databases that have been merged into others can be annotated with the ID of the database they have been merged into. This allows the retention of status and other annotations for the original database while providing a link to the new database entry. Updated annotations are then being tracked in the database entry of the database it was merged into. An example is the database "lncRNAdb" which is now available through "RNAcentral".

\subsection*{Website}
Both meta-registry websites have been adjusted and improved to reflect the unification, using a single, combined database and codebase. However, both websites remain independent regarding their database subsets of NAR catalog entries and miRNA databases. Future developments are directly available for both registries. Annotations for databases listed in both registries are shared, reducing the annotation effort as well.

The browsing page has transitioned from listing NAR catalog entries to database entries with their status, URLs, publications, and more. Detail pages for database entries list all available annotations as well as associated NAR catalog entries. Old links to detail pages are redirected to their new database entry equivalents. Where NAR catalog entries are associated with multiple database entries, the user is provided with a redirection choice. Annotation of entries by logged-in users is performed on the respective detail pages as before.

A new about page lists information on how to provide feedback, cite the meta-registries, data license, and the ability to download the latest registry annotations. In addition to the annotation download a new cross-references download has been added. This TSV file lists all database entries and their cross-references to the other meta-registries mentioned before. Other meta-registry maintainers are invited to use this cross-references file to update and extend their own information with the hope that they may provide such a file as well in the future for an improved exchange of information.

\subsection*{Statistics}
Existing statistics regarding the status, download availability, years of last database update, and curation staleness have been adjusted to use the new database entries without the need to de-duplicate NAR catalog entries. Furthermore, the addition of all database publications allows a miRNAverse statistic to be used for NAR db status as well, visualizing the publication and availability of databases over time. Two new statistics have been developed utilizing this data as well. First, a simple line plot visualizes the relative unavailability of databases over time. Second, a combined dot plot and heat map visualizes the comparison of potential and actual database longevity using the earliest annotated publication. The heat map element of the statistic shows the relative number of databases at the specific dot.


\section*{Results and Discussion}
The latest version of NAR db status increased the number of unique databases to 2,082, see Table~\ref{table1}, while miRNAverse currently lists 194 databases of which 55 overlap. The 2024 Nucleic Acids Research database issue lists more database additions \citep{nar2024}. However, as databases are mostly added to the NAR catalog on request, the actual number of databases differs. No new entries have yet been added from the recently published 2025 NAR database issue as of 30.01.2025. Following, the statistics of the NAR db status meta-registry are described in detail.

\begin{table}[!ht]
\centering
\caption{{\bf Comparison of NAR db status entries in 2023 and 2025 and miRNAverse}}
\begin{tabular}{|l|r|r|r|}
    \hline
                                   & \textbf{2023} & \textbf{2025} & \textbf{miRNAverse} \\ \hline
        Unique databases           & 2,025         & 2,082         & 194                 \\ \hline\hline
        Unavailable                &   466         &   587         & 76                  \\ \hline
        Recheck                    &   100         &   107         & 8                   \\ \hline
        Available                  & 1,416         & 1,349         & 107                 \\ \hline\hline
        Unknown updated            &   834         &   770         & 114                 \\ \hline
        Downloadable               & 1,011         & 1,078         & 80                  \\ \hline
    \end{tabular}
\label{table1}
\end{table}

\begin{figure}[H]
\centering
\includegraphics[width=\textwidth]{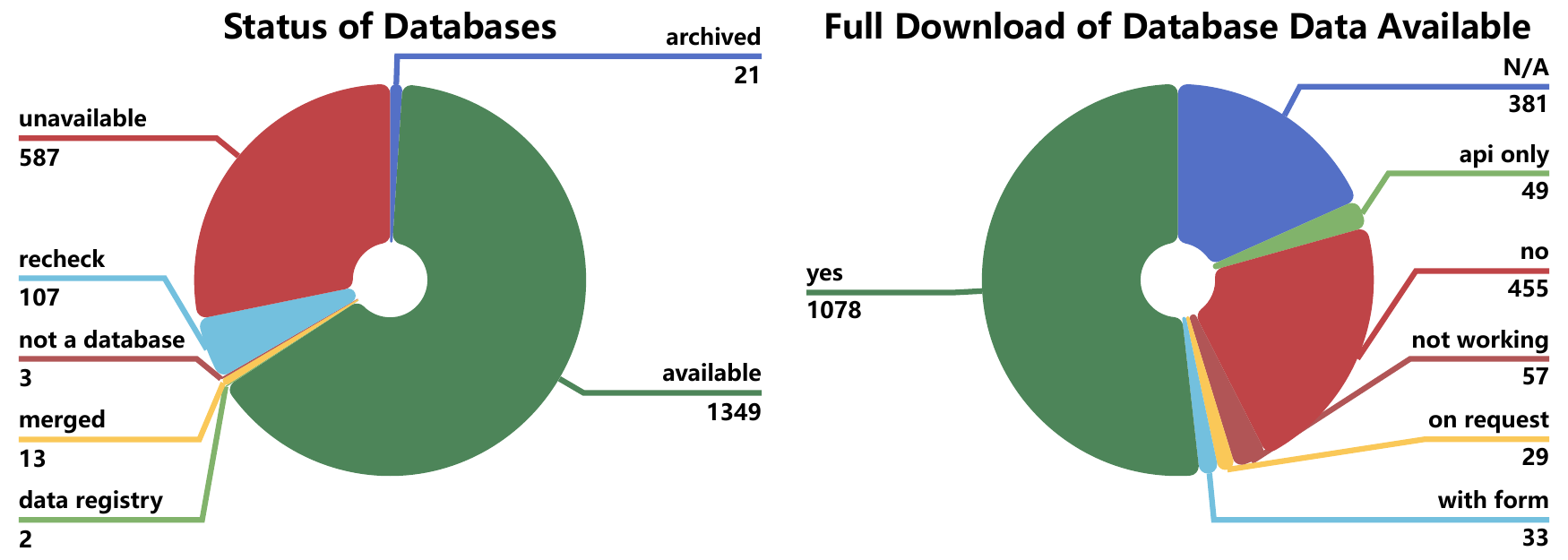}
\caption{{\bf Database status and download availability}
Two pie charts visualizing the distribution of database status and full database download availability.}
\label{fig1}
\end{figure}

Fig.~\ref{fig1} shows nearly two-thirds of databases are available. The number slightly decreased from 2023 while the number of unavailable databases increased by 119. This indicates more databases have disappeared than new ones have been added.

The number of downloadable databases increased compared to 2023 and is now over 50\%, see Fig.~\ref{fig1} on the right. The number of 'N/A' and 'no' decreased showing that some additional download options could be found and are now annotated as part of the other categories.

\begin{figure}[H]
\centering
\includegraphics[width=\textwidth]{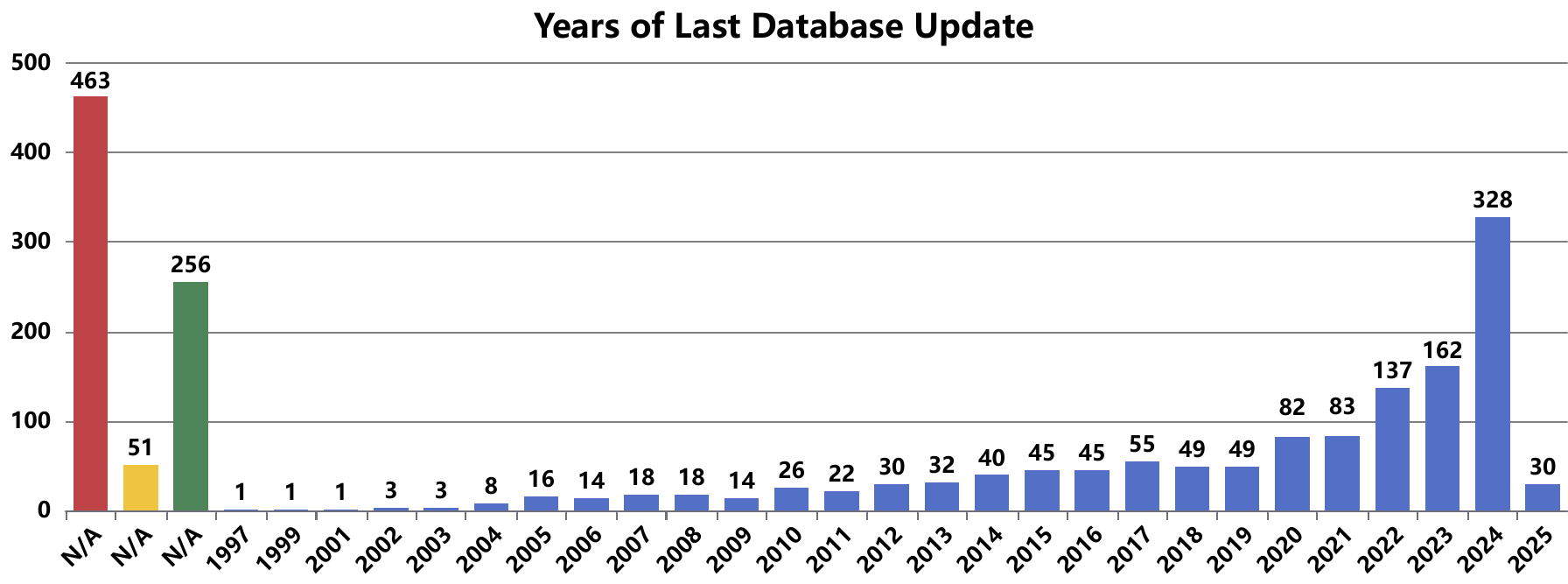}
\caption{{\bf Distribution of database updates}
Bar plot showing the latest year of database updates. Where no information exists the bar is split into three bars representing available, unavailable and other database statuses.}
\label{fig2}
\end{figure}

More databases have been updated in 2024 (328 plus 30 in 2025) compared to the 318 reported in 2023, see Fig.~\ref{fig2}. Table~\ref{table1} and Fig. \ref{fig2} show that the number of unavailable databases did increase. Approximately one-third of the databases in NAR db status have been updated in the last three years.

\begin{figure}[H]
\centering
\includegraphics[width=\textwidth]{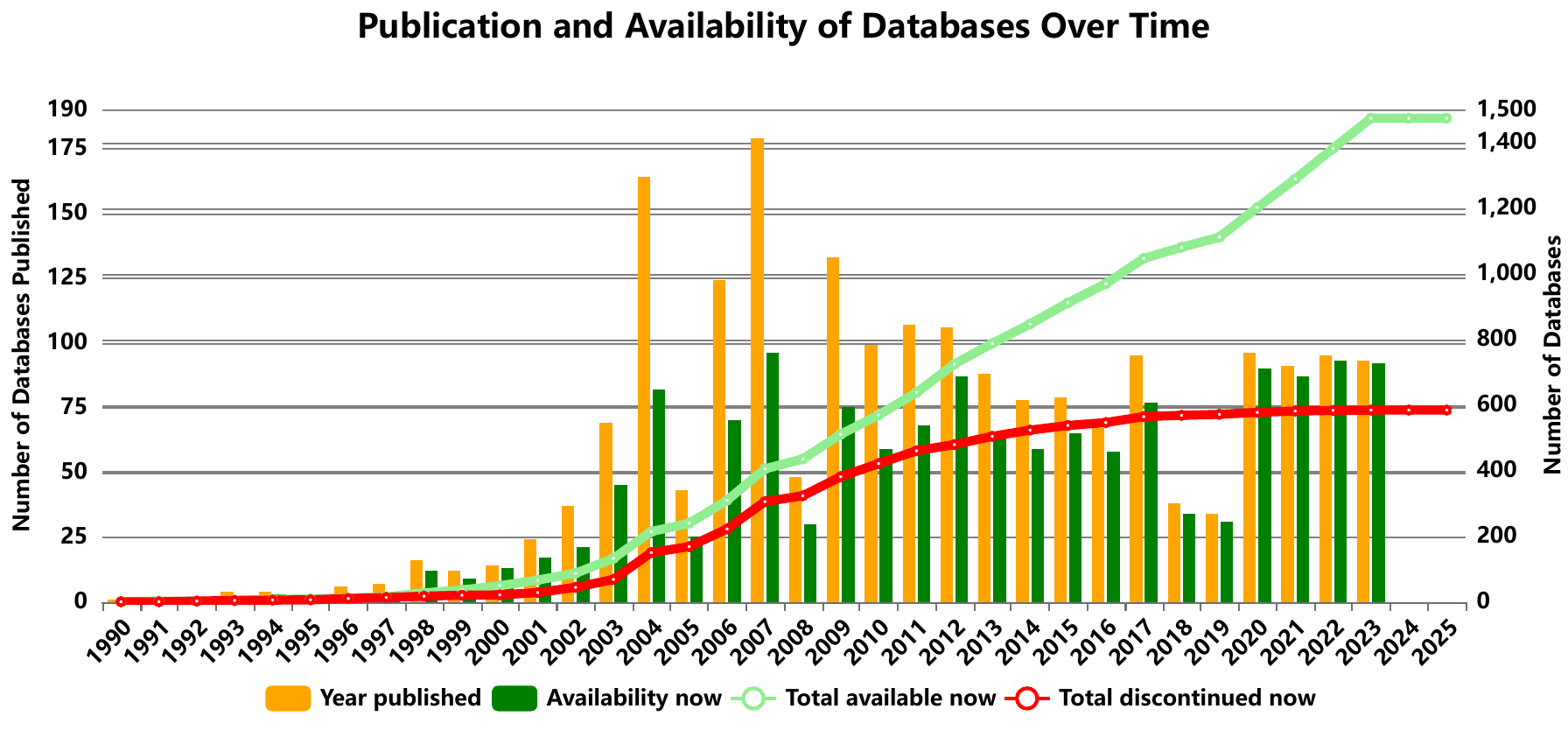}
\caption{{\bf Publication and availability of databases over time}
Bar plot visualizing the number of databases first published in the respective year (orange) and how many of them are still available now (green). Two line plots represent the cumulative numbers of available (green) and unavailable (red) databases.}
\label{fig3}
\end{figure}

The plot in Fig.~\ref{fig3}, previously only available in miRNAverse, shows the changing availability of databases over time relative to their first publication. This shows that 179 databases were first published in 2007 of which 95 are still available. Many older databases are still available with the oldest being the endogenous regulatory oligopeptide knowledgebase from 1990 \citep{oldestDB1990}.

\begin{figure}[H]
\centering
\includegraphics[width=0.65\textwidth]{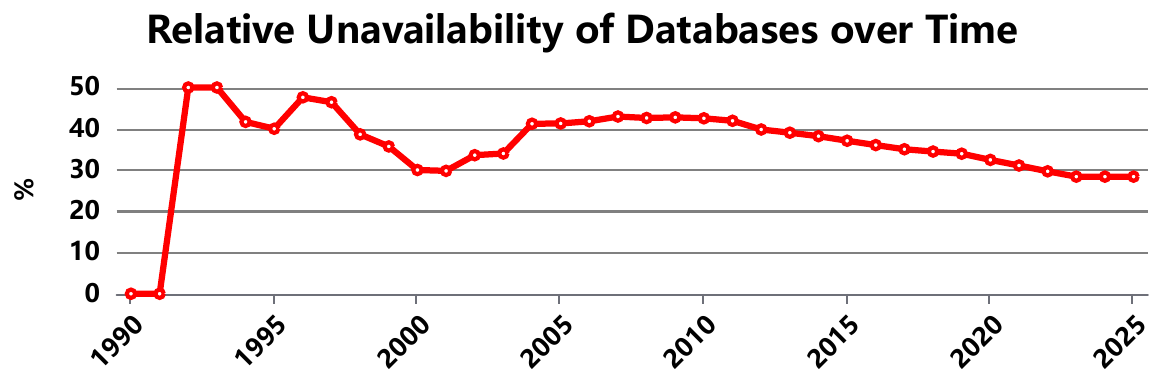}
\caption{{\bf Relative unavailability of databases over time}
Line plot visualizing the unavailability of databases over time relative to the total number of databases.}
\label{fig4}
\end{figure}

From 2018 to 2023 less than ten databases are unavailable per year. This agrees with the observation that new databases are more likely to be available for at least a few years. From 2002 to 2011 almost half of the databases are unavailable now. Overall, the cumulative number of available databases (green line in Fig.~\ref{fig3}) is always higher than the cumulative number of unavailable databases (red line). A different visualization of this observation is shown in Fig.~\ref{fig4} where the relative unavailability never passes the 50\% line. The cumulative numbers show a trend that the number of available and unavailable databases
approach each other as time passes. This trend diverges for more recent publication years, indicating a delay from publication to database disappearance.

\begin{figure}[H]
\centering
\includegraphics[width=\textwidth]{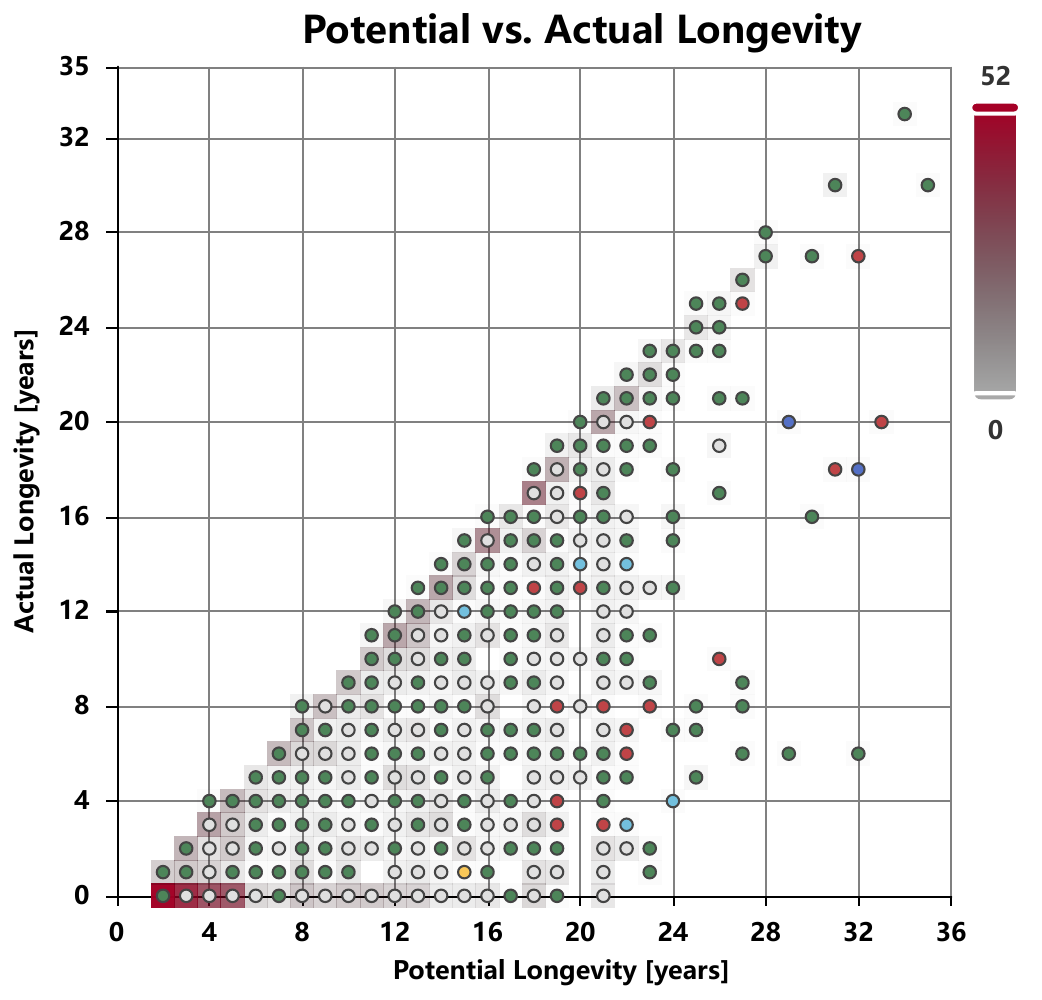}
\caption{{\bf Potential to actual longevity}
Scatter plot visualizing the potential longevity of a database on the x-axis and the actual longevity on the y-axis in relation to the oldest publication and last update year. Dots are colored similarly to the status categories from Fig.~\ref{fig1}. If not all databases represented by a dot have the same status the dot is drawn white. Additionally, dots have a heatmap background colored from transparent to red representing the number of databases per dot. A single database is shown with a transparent heatmap background and the maximum of 52 databases with a red background.}
\label{fig5}
\end{figure}

Fig.~\ref{fig5} visualizes the potential versus the actual longevity of databases relative to their oldest publication. All dots on the diagonal represent databases from different years that have been updated in 2025. The endogenous regulatory oligopeptide knowledgebase with a potential longevity of 35 and actual longevity of 30 years is the oldest database with publication and update annotations. With a potential of 34 and actual longevity of 33 years, PROSITE is the longest continuously updated resource with both annotations in the registry. The dots with highest heatmap values (greater than 30) are visualized in red and represent the most aggregate databases. These were published in recent years and have not been updated since.

\begin{figure}[H]
\centering
\includegraphics[width=\textwidth]{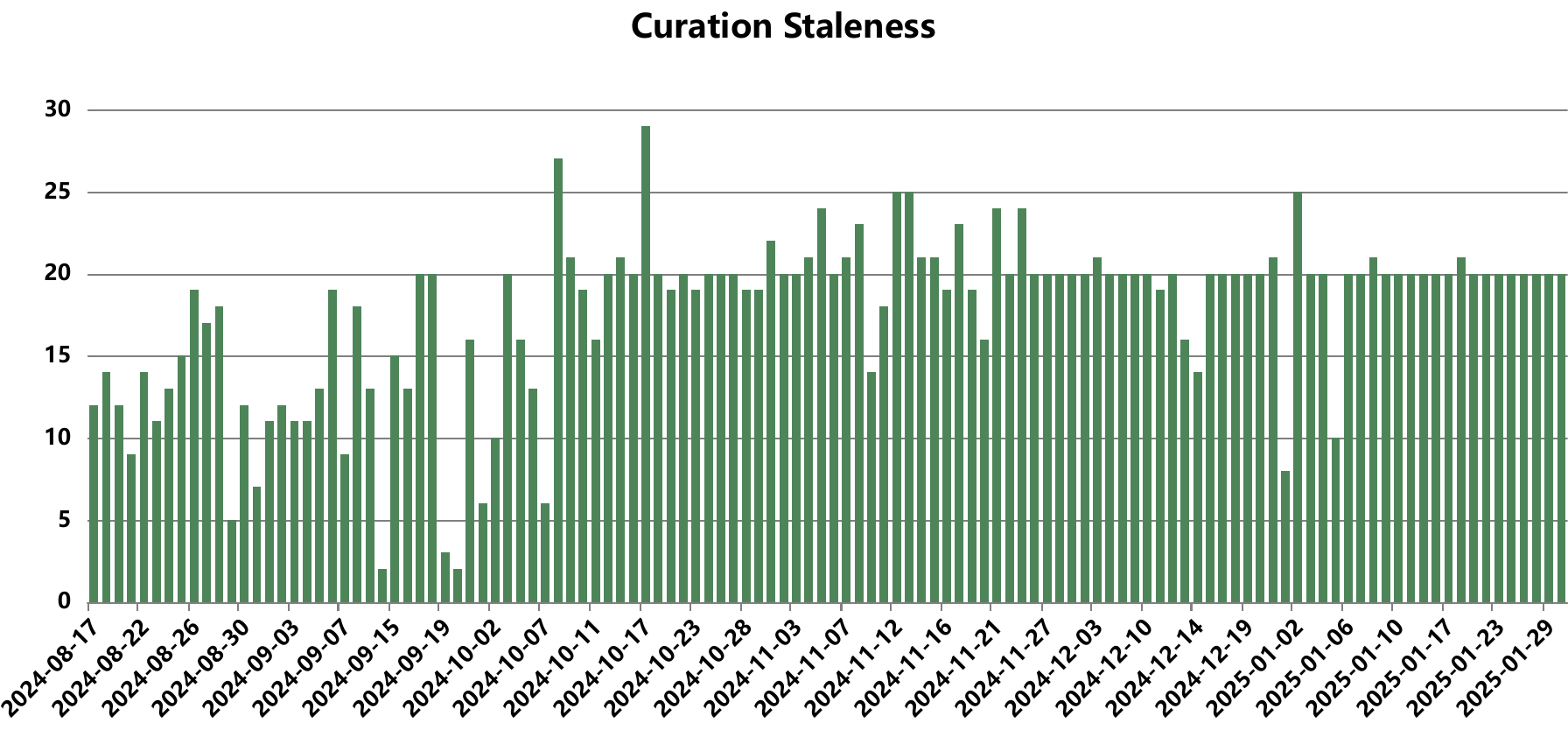}
\caption{{\bf Curation staleness}
Bar plot visualizing the curation staleness of all NAR db status entries. The average curation staleness based on the latest revisions is 78.71 days. Each database is usually updated three times a year.}
\label{fig6}
\end{figure}

The oldest annotation revision in the database is from June 2022 and the newest from January 2025, spanning two and a half years of continued manual curation. As visualized in Fig.~\ref{fig6}, the curation process and it's staleness are transparent. Currently, the average curation staleness based on latest revisions is 78.71 days. In our previous publication, the declared goal was the annotation of 20-40 entries per active day. Generally, this range is achieved. Depending on the individual situations of holidays, illness, or others of the curators the active days are separated by gaps, but usually consecutive and caught up with additional annotations if necessary. Overall, a steady and continuous annotations ensures the up-to-date annotation of all database entries.

\section*{Conclusion}
The unification of the meta-registries NAR db status and miRNAverse was successfully implemented, simplifying maintenance and future updates for both. Entries listed in both database subsets are now annotated only once reducing the workload.

The addition of publication annotations to NAR db status and annotation revisions to miRNAverse provide more in-depth statistics and insights into database longevity and availability. With over two years of continued database annotation, transparent revisions, and curation statistics, NAR db status has proven manageable in size for two active curators and is prepared for future NAR updates and database additions.

With all annotations free to download and the the new database cross-references download the hope is for other meta-registries to incorporate the provided information in the shared effort to support researchers in database selection. Both meta-registries will continue to provide up-to-date annotations. New developments are planned to further simplify the annotation process and new statistics are planned to find suitable databases with even more in-depth information. Feedback or requests for collaboration are always welcome and can be submitted as outlined on the meta-registries about page. NAR db status is now available at the new URL \url{https://nardbstatus.de} and miRNAverse at \url{https://mirnaverse.de}.


\bibliographystyle{unsrtnat}

\end{document}